# ALTERNATIVE REFLECTIONS ON GRAVITATION

JOHANN ALBERS

Fachbereich Physik der Universität des Saarlandes

66041 Saarbrücken, Germany

## ABSTRACT

It is assumed that the primary interaction between two masses $m_1$ and $m_2$ is not attractive as postulated by Newton's law of gravitation, but repulsive. Both $m_1$ and $m_2$ emit and absorb gravitational radiation. Corresponding to the laws of optics the absorption is connected with an impulse transfer that produces the repulsive force. If, however, $m_1$ and $m_2$ are embedded in the gravitational radiation produced by all the masses of the universe the absorption by $m_1$ and $m_2$ leads to a reduction of the intensity of the gravitational radiation between them, thus creating an attractive force exactly as described by Newton's law. The so called universal gravitational constant is no constant. It locally depends on the arrangement of the masses in the universe. It can accept high values which are usually explained by the existence of dark matter. Due to the primary forces of repulsive nature between all masses the expansion of the universe is an intrinsic property. The balance between primary, repulsive and secondary, attractive gravitational forces can stabilize highly concentrated mass accumulations as they are observed in globular clusters and the bulges of galaxies.

*Subject headings:* alternative theory: gravitation - Newton`s law of gravitation - dark matter - expanding universe - cosmology

## 1. INTRODUCTION

More than three centuries ago Isaac Newton formulated the law of gravitation that describes the mutual interaction between any two masses $m_1$ and $m_2$. The interaction is of attractive nature and can be described as a force F the absolute value of which is proportional to the product of both masses and inversely proportional to the square of its distance. The negative sign in equation (1) points to the attractive nature of F.

$$F = - G * m_1 * m_2 / r^2 \tag{1}$$

If this relation is valid for two mass points then it holds also for two extended spheres if the mass distribution of the two spheres is homogeneous and if r is the distance between their centres. For the proof of this statement which was already performed by Newton and which can be found in textbooks ( Falk & Ruppel 1989) it is sufficient that the mass distribution is homogeneous within the space between two concentric spheres.

The factor of proportionality G in equation (1) could be determined in laboratory experiments with masses $m_1$ and $m_2$ in the kg-range with increasing accuracy to about $G = 6.6726$ $Nm^2kg^{-2}$. The weight of a known mass then directly leads to the mass of the earth. Knowing the mass $m_1$ of the earth and assuming that the centrifugal force $F_z = m_1 * v^2 / r$ equals the gravitational attraction due to equation (1), one can deduce the mass $m_2$ of the much heavier sun.

$$m_2 = v^2 * r / G \qquad (2)$$

Equation (2) can also be used to describe the rotation of single stars around the centre of galaxies. If the distance r between the star and the centre and the speed of rotation v are known then there follows the value $m_2$ of all masses located within the sphere with the radius r. In the simplest case considered here spherical symmetry and homogeneous density are assumed in every space between two concentric spheres.

It should be mentioned that taking equation (2) as a basis one can derive the value of $m_2$ from measured data of v and r only if the constant G resulting from terrestrial experiments is assumed to be valid for the whole universe. That means that the value of G obtained from tests in the kg-range must be the same across about 45 powers of ten up to the interaction between clusters of galaxies with masses up to $10^{15}$ times the solar mass. That is probably not only in the opinion of Berry (1990) a colossal assumption. Because there is obviously no commonly accepted model from which another assumption can be plausibly derived, G is commonly assumed as a constant valid in the whole universe, i.e. G is considered as the "universal constant of gravitation".

## 2. COSMIC FINDINGS

By use of modern spectroscopic methods today information is obtainable from regions of the universe where the distances and mass concentrations exceed those of the solar system by many orders of magnitude. Since 1929 it has been known through the investigations of E.

Hubble that the universe is expanding. The big-bang theory is the favoured explanation (Berry 1990) of this expanding behaviour that is opposing the attractive gravitational effect.

From spectroscopic measurements on single stars within galaxies gravitational interactions are derived ( in analogy to the remarks to equation ( 2)) which lead to masses $m_2$ of the galaxies much higher than those expected on the basis of the well-known mass-luminosity relation. The commonly accepted explanation assumes the existence of the so called dark matter (Unsöld & Baschek 1991). The amount of dark matter is the higher the stronger the mass concentration is in the region under consideration. In the surrounding of small galaxies with about $10^7$ solar masses one does not need the dark matter to explain the gravitational effects. However, where super-clusters of galaxies with up to $10^{15}$ solar masses have concentrated the necessary amount of dark matter exceeds that of the luminous mass by a factor of about 100. If the logarithm of the dark matter is drawn in a diagram against the logarithm of the luminous matter (Kraus 1989) an almost linear behaviour results. It can be well described by an exponent of about 1/4 although there is a considerably high limit of error. When the measured mass-luminosity value is drawn against the diameter of the considered cluster in a double-logarithmic diagram (Rubin 1984) one finds a proportionality over a large range which results in a value of 1/3 for the above mentioned exponent, provided that the same mean mass density is assumed for all clusters.

It is an astonishing fact that on the one hand the amount of dark matter increases with increasing density of the luminous matter towards the centre of the superclusters of galaxies and that on the other hand the necessary amount of dark matter decreases in individual galaxies with increasing density of the luminous matter towards the core of the galaxy. Following D.W. Sciama et al. (1992) one might therefore think about two different types of dark matter. Considering the fact that in the literature there is a further distinction between hot and cold dark matter one can already speculate about four different types of dark matter. Aside from the gravitational effects for which the existence of the dark matter was postulated, there is, however, no other experimental proof of its presence. For that reason it is understandable that M. Berry (1990) states that the doubts about its existence increase.

Also, it is understandable that with regard to several not convincingly explained cosmic effects in astrophysics and due to the respect for the "exact laboratory physics" in this point one would rather prefer to waive the well established mass-luminosity relation then calling in question Newton's law of gravitation with its universal constant which was confirmed by laboratory physics.

A striking cosmic finding is the existence of an astronomical high number of about $10^{11}$ galaxies in the universe that themselves consist of about $10^{11}$ stars (Berry 1990). Our actual observations provide insights in states of these galaxies which, depending on their distance, still exist or existed a long time ago. Considering the similarity, especially that of the cores of near and distant galaxies, the question arises whether these highly compressed arrangements of stars should not represent more or less stable final states.

Another striking finding is the existence of more than 100 globular clusters just in our own galaxy in which old stars of population II have assembled similar to the cores of galaxies with more and more increasing density towards the centre. A typical globular cluster contains several hundred thousand stars in a region with a diameter of about 40 pc (Unsöld & Baschek 1991). If one assumes the mass of our sun with $m_0=2.0*10^{30}$ kg as the mean mass of a star then a mean density of $2.0*10^{-19}$ kg/m$^3$ results for a globular cluster with $10^5$ stars. Compared to the mean density of the luminous matter in the universe of $5*10^{-28}$ kg/m$^3$ ( all constants used here are taken from Unsöld & Baschek (1991) if given there ) this implies an immense compression by a factor $4.0*10^8$ in these small volumes of globular clusters. As with the cores of galaxies that may be seen as huge globular clusters (Asimov 1979) one may ask again why we can see so many systems in which enormous masses have concentrated almost to a point, but even after billions of years have not concentrated exactly to a point.

The law of gravitation that is naturally seen in connection with concepts like gravitational collapse and dark holes is really not very suitable for the explanation of these states as almost stable final states. Apart from the fact that the findings mentioned here are needed for a comparison with the ideas to be developed in the following, they should also show to what small an extent single hypotheses are commonly accepted by astrophysicists.

## MODEL HYPOTHESES

In order to elucidate the model hypotheses to be developed in the following a paradigm belonging to the field of mechanics may be mentioned here which can contribute to the understanding due to its strong analogy.

When O. von Guericke evacuated the volume inside his Magdeburg hemispheres he produced an attractive, directional interaction between the two hemispheres that was directed along the axis of symmetry of the system. It is well known that this was not a discrete and unexplainable effect. Rather, this attractive interaction is created as a secondary effect by the shielding of the volume inside the two hemispheres against the hydrostatic atmospheric pressure. The primary effect, however, is the repulsive interaction between the gas molecules that results from the impulse transfer and produces the hydrostatic pressure.

In analogy to this effect the following is postulated concerning the gravitational effect:

***The primary interaction between masses is not of attractive but of repulsive nature.***

With regard to its sign the interaction postulated this way is, contrary to the attractive interaction according to equation (2), in accordance with the electromagnetic interaction, where as well with the magnetic as with the electrical interaction sources of the same kind (positive or negative electrical charges or magnetic poles) are repelling one another, whereas sources of different kind are attracting one another. In the following a justification of this postulate will be constructed on the basis of plausible assumptions.

Equation (1) can be understood such that the effect of the mass $m_1$ results from a vectorial gravitational field ***E´*** with the value E´:

$$E´ = - G * m_1 / r^2. \tag{3}$$

The negative sign again indicates the attractive nature. E´ acts on the mass $m_2$ with a force ***F*** which is proportional to E´ and has the value $F = m_2 * E´$. For the following considerations, however, it is assumed that the effect produced by any mass m is not of vectorial kind but has properties that are similar to those of radiation. Therefore, in the following the notion gravitational radiation ( GR ) will be used, especially to differentiate between this and the usual approach to gravitation. However, that does not have to mean that the GR is really a type of electromagnetic radiation. This should be proposed analogously with respect to other notions known from the region of optics as intensity, emission, absorption, radiation pressure and so on. The use of these notions should serve primarily to elucidate that the following considerations are derived based on well-known physical effects and laws.

It should be mentioned that the existence of gravitational radiation was already predicted by Einstein in the scope of his general theory of relativity. Between 1940 and about 1970 many groups were concerned with the elaboration of the theory of the emission and absorption of gravitational radiation and especially of the retroaction on the emitting system (Danzmann & Ruder 1993).

It will be assumed that the interaction between the GR emitted by a mass $m_1$ and a mass $m_2$ is generated by the fact that the GR emitted by $m_1$ is partially absorbed by $m_2$, in the course of which a repulsive force is produced. The same holds for the GR emitted by $m_2$ and absorbed by $m_1$. This is in complete analogy to the optical radiation where the impulse transfer connected with the absorption ( or reflection ) of photons leads to a repulsive force acting on the absorbing medium. With regard to these considerations some more details are assumed as plausible as possible even if no direct proofs of these assumptions exist.

Regarding the emission of GR it can hardly be assumed that its emission by masses can be described in the simplest way by a linear relation independent of the mass distribution with densities reaching from values below the mean density of the universe of $5 * 10^{-28}$ kg/m$^3$ to the more than 30 orders of magnitude higher densities in the stars. If in spite of this the emission of GR is assumed to be describable by the product $E * m$, i.e. the GR is proportional to m with E as a factor of proportionality, semiquantitative conclusions can be drawn and comparisons can be performed only if the spatial distributions of masses have similar structure.

In analogy to optics it is assumed that the intensity of the GR emitted by different masses can be summed up linearly as it is the case with incoherent radiators and that the intensity decreases inversely to the square of the distance. The repulsive force F acting from $m_1$ on $m_2$, and vice versa, can be described by

$$F = E*m_1/r^2 * A*m_2 * P \qquad (4)$$

The term $E*m_1/r^2$ represents the intensity I of the GR produced by $m_1$ that is valid in a distance r at the place of $m_2$. $A*m_2$ represents the absorption of the GR by $m_2$ that is assumed to be linear in $m_2$. P symbolizes the factor of proportionality between the absorbed GR and the repulsing force resulting from the impulse transfer of the absorbed GR. The correct dimensioning of the effect described by the symbols I, E, A ,P will not be discussed here in more detail. Equation (4) describes the above postulated primary interaction between two masses $m_1$ and $m_2$. This interaction is not attractive but repulsive.

If one considers not only two isolated masses $m_1$ and $m_2$ but aggregations of masses then the intensity I of the GR at a certain reference point Q is achieved by adding the GR of all masses $m_i$ surrounding Q at a distance $R_i$. Instead of the correct integral a summation may be used here.

$$I = \text{Sum}_i ( E * m_i / R_i^2) \qquad (5)$$

As already mentioned above the masses are seen as incoherent radiators for which the intensities rather than the amplitudes have to be summed up and where the usual $1/R^2$ dependence is assumed. Proposing a central-symmetrical mass distribution around Q the addition of a vectorial gravitational field according to equation (3) yields the value zero of the effective gravitational field at the reference point Q. The intensity I according to equation (5), however, has a non-zero value at Q that may be compared with the above mentioned hydrostatic pressure.

If one considers mass aggregations with spherical symmetry, homogeneous density, a radius R, and an overall mass M then the summation of $m_i/R_i^2$ in equation (5) will give a value $3*M/R^2$ and an intensity $I(R) = E*3*M/R^2$ at a point Q in the centre of the sphere. If Q migrates away from the centre then this value and the corresponding intensity are reduced. With Q at the periphery of the sphere the value of I reduces to 50% of the value at the centre. For distances r>>R, I can be increasingly well described by $I(r) = E*M/r^2$. If Q shifts away from the centre the intensity I gets a directional component directed from the centre towards Q ( here, the integration is performed only with the cosine-component effective along this direction, similar to the integration in equation (1)). This directional component $I_d$ of I has a zero value at the centre ( equal radiation intensity from all directions ) and increases linearly with the distance from the centre to

$$I_d = E *M/R^2 \qquad (6)$$

at the periphery. This value is 1/3 of I(R) at the centre. At distances r with r > R the intensity $I_d(r)$ reduces exactly proportionally to $1/r^2$.

A simple and illustrative way to compare the intensities of mass aggregations with spherical symmetry but with different size and density is performed by dividing the sphere into calottes with equal thicknesses. Every calotte contributes the same amount of intensity according to the term $m_i/R_i^2$ in equation (5). Therefore, the ratio of the intensities at the centre of two spheres with radii $R_1$ and $R_2$ and densities $d_1$ and $d_2$ will be:

$$I_1/I_2 = d_1*R_1 / ( d_2*R_2). \qquad (7)$$

## 4. GRAVITATION AS A SECONDARY EFFECT

As a first example, a mass $m_1$ is considered which is radiation-exposed from all surrounding masses in the universe with equal intensity of the GR from all directions. For simplicity it is assumed that the masses are distributed with spherical symmetry in the universe. The forces produced in $m_1$ by the absorption of GR are vectorially summed up to zero, exactly as with the conventional treatment of the vectorial gravitational fields of these masses according to equation (3). Analogous results are valid for a mass $m_2$.

In a second example, two masses $m_1$ and $m_2$ are considered. For simplicity, $m_2$ is assumed to be a small, punctual mass and $m_1$, seen from $m_2$, is a small part of a thin calotte with a solid angle $w_1 \ll 1$. The spatial extension of both masses is assumed to be small compared to its distance. The GR penetrating $m_1$ in the direction towards $m_2$ is attenuated proportionally to $m_1$ by the part $A*m_1$ ( factor $A*m_1$ ). From this part the absorption by $m_2$ is missing ( factor $A*m_2$). Therefore, there is also missing the repulsive force produced by the impulse transfer (factor P). The overall effect is an attractive force between $m_1$ and $m_2$. The attenuation of the GR is, as long as it can be considered to be very small, obviously proportional to $m_1$, no matter whether a change of $m_1$ results from an increase in the thickness of the calotte or from an increase of the differential solid angle. Obviously there exists also a proportionality to two other factors. These factors are $m_2$ and the reciprocal of the square of the distance, $1/r^2$. Therefore, it follows: The attractive force between two mass points $m_1$ and $m_2$ which are small compared to their distance is proportional to the product of these masses and inversely proportional to the square of their distance. Taking into account Newton's proof mentioned above this holds also for two extended masses $m_1$ and $m_2$ with homogeneous density. It follows

$$F = -I_0 * A*m_1 * A*m_2 * P * 1/r^2 \qquad (8)$$

for the attractive force between $m_1$ and $m_2$. Equation (8) can be written in analogy to equation (1) in the form

$$F = - G_f * m_1*m_2 / r^2 \qquad (9)$$

where the factor of proportionality $G_f$, denoted as gravitational factor in the following, is given by

$$G_f = I_0 * A * A * P. \qquad 10)$$

$I_0$ is derived from equation (5) where the sum extends over all masses of the universe with their real spatial distribution.

The absorption of the GR symbolised by the factor A was assumed until now to be so week that the exponential law valid for absorption effects could be applied in its linearized form. Concerning this point numerical calculations using the exact exponential form of absorption were performed on a PC with the following results: With stronger absorption the $1/r^2$ dependence according to equation (9) has to be corrected. With analytical corrections fitted to the numerical values of the r-dependence orbital motions were calculated numerically and extrapolated to the orbital data of Mercury on its orbit around the sun. An absorption strength by which a beam penetrating the centre of the sun is weakened by 0.8 % would lead to a advance of Mercury's perihelion by 43" per century, just the value of the anomalous precession of the perihelion (Berry 1990) which is already explained on the basis of the general theory of relativity. If the precession of Mercury's perihelion due to the absorption of GR by the sun has to be far smaller than this value then the absorption of GR even by the sun must be assumed to be small enough to allow the linearization of the law of absorption in agreement with the above used assumption. As this small absorption of the GR enters quadratically into gravitation according to equation (8) it may become understandable why for example the gravitational interaction between a proton and an electron is so small compared to the electrostatic one.

When here in connection with the laws of electromagnetic radiation values of absorption are discussed which even after penetrating the sun are extremely small then at first sight this may seem unrealistic. It may, however, with regard to a hypothetical candidate for this radiation, be pointed to the neutrino. Depending on their properties neutrinos must be treated as extremely relativistic particles (Falk & Ruppel 1989). Thus they may in principle be identified with GR. For neutrinos it is commonly assumed that their absorption by masses, even when penetrating the sun, is extremely small.

The law of gravitation in equation (9) must therefore be considered as a limiting case for sufficiently small masses $m_1$ and $m_2$ and it is, as well as the linearized absorption described by the factor A, not valid for extremely large and dense mass conglomerations. The primary, repulsive interaction between two masses $m_1$ and $m_2$ according to equation (4) must of course be added to the attractive interaction according to equation (8). This is possible because both equations have the same $1/r^2$ dependence. When the gravitational factor $G_f$ is defined in equation (9) this contribution should be considered as being included. The effectively occurring attractive interaction is possible only if the attenuation through $m_1$ of the GR produced by all the masses in the universe ( factor $I_0*A*m_1$ in equation (8) exceeds the GR which is produced by $m_1$ itself ( factor $E*m_1$ in equation (4) ). And this case is assumed for all masses $m_1$ and $m_2$ in our solar system.

The most essential statement of these considerations can be expressed as follows:

***The attractive gravitational interaction between two masses $m_1$ and $m_2$ is a secondary effect that results from their mutual screening against gravitational radiation produced by all masses in the universe.***

## 5. CONSEQUENCES OF THE MODEL

At first sight there seems to be no advantage with the derivation of equation (9) on the basis of effects and concepts well known in optics as a possible foundation of the law of gravitation given by equation (1) especially if the reflections are restricted to this law of gravitation. At second sight, however, there appear some differences which result from the fact that equation (9) is derived from model considerations from which more conclusions can be drawn.

Because the primary interaction between all masses in the universe is of repulsive nature the expansion of the universe follows as an intrinsic property as with a cloud of gas that is not encased. Therefore, there exists no problem with the flatness of the universe for which otherwise the overall mass of the universe is a critical parameter. The expansion of the universe must exist independently of the explosive nature of the big bang.

In contrary to the constancy in space and time of the universal gravitational constant in equation (1) the gravitational factor in equations (9) and (10) is, due to $I_0=I$ in equation (5), dependent on the distribution of masses around the point of reference. What are the consequences that must be expected?

As a first case, the influence of the inhomogeneous distribution of mass in the universe may be considered. This inhomogeneous distribution is of course also determined by dynamical effects, but this will not be considered here in more detail. Our galaxy is located within the local group of galaxies (Cambridge Enzyklopädie der Astronomie 1989) which does not represent a considerable increase in the mean density of mass and in which there is almost no dark matter necessary to explain the gravitational effects. However, that these statements are very uncertain may be inferred from the fact that according to Zaritzky (1992) even recent statements about the necessary amount of dark matter in our galaxy differ by a factor of about 6. Due to the relatively small densification of mass it is assumed that the gravitational constant in equation (1) that is valid in our solar system corresponds to a gravitational factor $G_s$ ( s for solar system ) which is not significantly different from the value of $G_f$ in equation (10). $G_f$ is determined by the intensity $I_0$ which according to equation (5) is produced by the roughly uniformly distributed overall mass in the universe.

Displaying the number of galaxies in a cluster of galaxies against the diameter of the clusters one finds for the largest clusters a number of about 2000 galaxies within a diameter of 1 Mpc (Cambridge Enzyklopädie der Astronomie 1989). With the commonly assumed number of about $10^{11}$ stars per galaxy and a star mass according to that of our sun there results a mean density $d_1 = 2.6*10^{-23}$ kg/m$^3$ with a radius $r_1 = 0.5$ Mpc in the cluster. If one takes in equation (7) the mean density of the universe, $d_0 = 5*10^{-28}$ kg/m$^3$ as $d_2$, and the radius $r_0=13$ billion light-years ( resulting from Hubble´s constant with a value of 75 km/s/Mpc ) as $r_2$ one gets $I/I_O = 6.5$. This means that in the centre of the cluster there exists a gravitational factor $G_c = (1+6.5) * G_s$. If one assumes, as discussed above, that $G_s$ has about the same value as the universal constant of gravitation in equation (1) then from our terrestrial view there exist strongly increased gravitational effects in the cluster. According to this, the measured increased gravitational effects seem to be really reasonable. They should therefore not be explained by the usually postulated dark matter where in this example the dark matter would exceed the luminous matter by a factor 6.5. Rather, it should be explained by a gravitational constant that is not universal but adapted to the gravitational factor $G_f$. Here, not so much importance should be attached to quantitatively correct values. For more quantitative calculations it must be recognized that the density within the cluster is not constant but shows increased values towards the centre thus increasing the value of $G_f$. A further point is the uncertainty in the mean mass density in the universe. Simply spoken, the main result is that due to the spatial concentration the above mentioned 2000 galaxies determine the gravitational effects in the centre of the cluster much more than the substantially higher number of $10^{11}$ galaxies in the universe.

In this model one may vary the overall mass of a cluster keeping the density constant. Regarding that the radius of the condensed region increases with the third root of the mass it follows with $I_2= I_0$ from equations (6) and (9) that the correlation between $G_f$ and this mass can be described by a slope 1/3 in a double logarithmic representation. With the equivalence of $G_f$ and the dark matter this simplified model corresponds well with the experimental results mentioned in chapter 2.

As a second case, the competition between the primary, repulsive and the secondary, gravitationally attracting interaction will be examined. The gravitationally attracting forces can concentrate masses, however, not to any high value because with more and more increasing mass concentration the primary, repulsive effect will predominate and prevent further concentration. If one again inserts the radius and the density of the universe as $R_2$ and $d_2$ in equation (7) and takes the radius of 40 pc of a typical globular cluster as $R_1$ and its mean density which exceeds that of the universe by a factor of $4*10^8$ as $d_1$ one gets the ratio $I_1/I_2=2$. That means that the masses of the globular cluster produce at its centre a contribution to the GR that is higher than that produced by all other masses of the universe. Of more importance, however, is the fact that the repulsive component $I_d$, which according to the calculation in chapter 3 ( equation. (6)) amounts to 1/3 of this value, is of the same order of magnitude as the GR that is produced by all masses of the universe and generates the attractive interaction. That implies that the gravitational attraction towards the centre of masses lying at the periphery of the globular cluster is neutralized to zero. Globular clusters and also the bulges of galaxies, appear therefore as already very stable final states of the effect of concentration of masses. Due to the missing attractive interaction no further mass wants to be incorporated. If one assumes that at every point inside the globular cluster the repulsive component has the same value then, according to equation (6), the density must increase towards the centre proportional to $1/r$. If the ability of masses to emit GR is already reduced due to the high density in the core or if the contribution of the outer part of the cluster to the attractive effect is taken into account then an even steeper increase of the density must be expected. However, instead of attributing too great an importance to the correct power law of the increase it is only pointed to the existence of the density increase towards the centre of globular clusters and the cores of galaxies occurring as well in reality as in this model.

Outside the core of galaxies the repulsive component $I_d$ of the GR decreases monotonically as explained in connection with equation (6). Therefore, stars rotating around the core far away in the spiral arms are aware of an effective gravitational constant according to the gravitational factor $G_f$ valid in this region. If the galaxy under consideration is situated in an extended region of highly increased density, as for example in the above mentioned accumulation of $10^{15}$ sun masses, there must be measured gravitational forces connected with theses stars which are far higher than the values calculated on the basis of the gravitational constant $G_s$ valid in our solar system. If one, however, explains these effects assuming dark matter then the curious behaviour of dark matter depicted in chapter 2 results. That includes an increase of the amount of dark matter correlated with the increased density of luminous matter towards the centre of extended regions with increased density but a decrease of the required dark matter correlated with the increased density of luminous matter towards the centre of single galaxies.

The most important results of the considerations in this chapter are summarized as follows:

***The gravitational forces cannot be described by a "universal" gravitational constant valid in the whole universe. The gravitational factor that has to be used instead of it is dependent on the mass accumulation of the universe seen from the reference point under consideration.***

## 6. CONCLUSIONS

Recent results in astrophysics can hardly be explained on the basis of Newton's laws, even by use of many additional ad hoc-assumptions adjusted to the different effects to be explained. Therefore, one must accept as a consequence, as done for example by H.J. Fahr (1992), that Newton's laws must be changed. And according to his opinion the focal point is which of his laws has to be replaced by a new one. If one uses the considerations, explained above and compared in several points with experimental results, one gets a basis on which Newton's laws really seem to be correct in the region of our solar system, and on which also experimental findings can be explained that result from far away regions of the universe and that considerably deviate from Newton's gravitational law.

The ideas described above were compared with the experimental facts using extremely simplified examples. However, only numerical calculations with a fit to the real mass distribution in the universe can provide information to clarify how the ability of mass to emit GR depends on mass. The above assumption of a simple linear dependence of the intensity of the GR on mass is surely not valid for a density like that of our sun that exceeds the mean density of the universe by about 30 orders of magnitude. This follows already from the fact that the sun is able to bind its planets, contrary to calculations after equation (6) with a linear dependence.

The above reflections were developed with strong reference to well-known notions and laws of optics. By use of these well-defined notions these model considerations may seem to be defined to sharply in some aspects. They, however, should by considered rather as one of many possible similar models. The validity of the gravitational law in equation (9) is not depending on the process by which the intensity $I_0$ of the GR in equation (10) is created.

It follows from Einstein's general theory of relativity that gravitational waves can be created only by temporal variations of quadrupolar mass distributions or higher multipole components ( Falk & Ruppel 1989). Whether there might have resulted quite different conclusions if the primary, repulsive interaction instead of Newton's gravitational law had been introduced competently into the general theory of relativity, can not be judged at this time. In any case it seems that especially the incorporation of Mach´s principle would not have raised severe problems.

In connection with the creation of GR the above mentioned example of the neutrino may be considered here again: Because neutrinos are produced especially in the interior of hot stars in such a case the GR would of course be produced by mass, however, not with the above assumed simplified linear dependence but with a dependence that is especially determined by the temperature of mass. As already mentioned above, comparisons between the amount of GR produced in different regions as universe, galaxies, and globular clusters can be drawn reasonably only if these regions are comparable concerning the relevant parameters ( here mass and temperature ).

If the alternative view presented above is seen as a step into a new direction, the last points clearly show that this direction cannot be defined very sharply at this time. Nevertheless further steps into this direction seem reasonable, especially when regarding the positive results of the qualitative comparison with experimental facts. At the end it may be pointed to an experiment which may be helpful for the decision whether the attractive interaction between two masses $m_1$ and $m_2$ according to equation (1) is only due to the existence of these two masses or whether the origin of the gravitational effects must be located in the width of the universe: the observation of a Foucault pendulum at the North Pole.